\documentclass[twocolumn]{aastex631}

\usepackage{amsmath}
\usepackage{subfigure}
\usepackage{bm}
\definecolor{blue-violet}{rgb}{0.30, 0.1, 0.89}

\begin{document}

\title{X-ray Polarization of the Intrabinary Shock in Redback Pulsar J1723$-$2837}

\author[0000-0002-9545-7286]{Andrew G. Sullivan}
\affiliation{Kavli Institute for Particle Astrophysics and Cosmology, Department of Physics, Stanford University, Stanford, CA 94305, USA}

\author[0000-0002-6401-778X]{Jack T. Dinsmore}
\affiliation{Kavli Institute for Particle Astrophysics and Cosmology, Department of Physics, Stanford University, Stanford, CA 94305, USA}

\author[0000-0001-6711-3286]{Roger W. Romani}
\affiliation{Kavli Institute for Particle Astrophysics and Cosmology, Department of Physics,
Stanford University, Stanford, CA 94305, USA}



\begin{abstract}
The intrabinary shocks (IBS) in spider pulsars emit non-thermal synchrotron X-rays from accelerated electrons and positrons in the shocked pulsar wind, likely energized by magnetic reconnection. The double-peaked X-ray light curves from these shocks have been well characterized in several spider systems. In this paper, we analyze Imaging X-ray Polarimetry Explorer ({\it IXPE}) observations of the redback pulsar J1723$-$2837 to examine the expected synchrotron polarization. Using advanced extraction methods that include spatial, temporal, and particle background weights, we constrain the polarization of the IBS. We compare different models for the magnetic field in the radiation zone and find that the best fit prefers a striped pulsar wind model {over other polarized models}, with maximum polarization degree of the IBS emission component $\Pi_{\rm IBS}=36^{+16}_{-15}\%$, in addition to an unpolarized non-IBS component. Since this is only 2.4$\sigma$, we cannot claim strong preference over an unpolarized model; we report a 99\% {confidence level} upper limit on the total polarization of both IBS and non-IBS components $\Pi_{99}<36\%$, {which is improved over the $50\%$ limit obtained in previous work}. The best-fit polarization of the IBS component is consistent with numerical simulations. Detailed tests of such models are accessible to future measurements. 
\end{abstract}


\section{Introduction} \label{sec:intro}

Spider pulsars are millisecond pulsars (MSPs) accompanied by low-mass companions in tight orbits with periods $P_b\lesssim1$ day.  Spider pulsars are often classed as black widows \citep{1988Natur.333..237F}, with sub-stellar $M_c\ll 0.1$ M$_\odot$ companions, or as redbacks \citep{2013IAUS..291..127R}, with $M_c\approx 0.1-0.5$ M$_\odot$. In these systems, the pulsar irradiates the companion and drives a stellar wind \citep{1988Natur.334..225K,1988Natur.334..684V}.
The relativistic pulsar wind and massive companion wind collide, forming an intrabinary shock (IBS), which dominates the X-ray flux in these systems \citep{2016ApJ...828....7R, 2017ApJ...839...80W, 2018ApJ...869..120W, 2019ApJ...879...73K}. In redbacks, the companion wind momentum typically dominates that of the pulsar, causing the IBS to wrap around the pulsar, while in black widows, the IBS wraps around the companion \citep{2016ApJ...828....7R}.

The pulsar wind is strongly magnetized, so the shock-accelerated particles emit prominent synchrotron X-rays in the post-shock flow \citep{1993ApJ...403..249A, 2017ApJ...839...80W, 2019ApJ...879...73K, 2020ApJ...904...91V, 2021ApJ...917L..13K}.
This flow accelerates to mildly relativistic speeds, so that the orbitally modulated IBS light curves often display two caustic peaks,  associated with beamed emission from relativistic particles traveling along the shock surface \citep{2019ApJ...879...73K, 2021ApJ...917L..13K, 2024ApJ...974..315S}. Shock-driven magnetic reconnection of the striped pulsar wind likely energizes the non-thermal particle population which emits the X-rays \citep{2012ApJ...756...33C, 2014ApJ...793L..20R, 2014ApJ...795...72L, 2019ApJ...879...73K, 2011ApJ...741...39S, 2021ApJ...908..147L,  2022ApJ...933..140C, 2024MNRAS.534.2551C, 2025MNRAS.tmp..266C, 2024ApJ...974..315S, 2025ApJ...984..146S, 2025ApJ...991...98S}.

The pulsar wind is expected to have a highly ordered toroidal magnetic field geometry \citep[e.g.][]{1999A&A...349.1017B, 2006ApJ...648L..51S}, with the field dissipating from the near zone to the termination shock. {\it IXPE} observations of pulsar wind nebulae (PWNe) confirm this picture, measuring very high $\gtrsim40\%$ synchrotron polarization and toroidal PWN symmetry \citep{2022Natur.612..658X, 2023NatAs...7..602B, 2024ApJ...973..172W}. In spiders, the pulsar wind termination shock is $10^6$ times closer to the pulsar than in typical PWNe, suggesting that the shock forms in even stronger and potentially more ordered fields. The partly annihilated striped wind should imprint its residual magnetic field geometry on the IBS, which can be probed by X-ray polarization studies \citep{2023IBSPolarization, 2025ApJ...991...98S, 2025arXiv250905240N}.

In this paper, we perform an {\it IXPE} polarization analysis of the redback J1723$-$2837 (J1723 hereafter). J1723 is a radio pulsar with spin period $P=1.86$ ms and spin-down luminosity ${\dot E}_{\rm PSR} = 4.7 \times 10^{34}$ erg s$^{-1}$. The pulsar has a $0.4-0.7$ M$_\odot$ companion orbiting with $P_b=14.8$ hr \citep{2013ApJ...776...20C}. Analyses of the X-ray emission strongly support IBS-domination \citep{2014ApJ...781....6B, 2017ApJ...839..130K, 2025ApJ...984..146S}. With one of the brightest IBSs known, J1723 is an ideal candidate to test IBS polarization models. {\citet{2025arXiv250905240N} have also performed an {\it IXPE} analysis of J1723, but only obtain a $99\%$ confidence upper limit on polarization of $\Pi_{99}\lesssim50\%$, using basic methods. Here we apply advanced data analysis methods to make more constraining measurements.} In Sec.~\ref{sec:observations}, we describe the {\it IXPE} observations and analysis methods used. In Sec.~\ref{sec:Model}, we compare proposed models with the {\it IXPE} data. We discuss our results and conclude in Sec.~\ref{sec:conclusion}.

\section{{\it IXPE} Observations}
\label{sec:observations}
\begin{figure*}
   \centering
   \includegraphics[width=\linewidth]{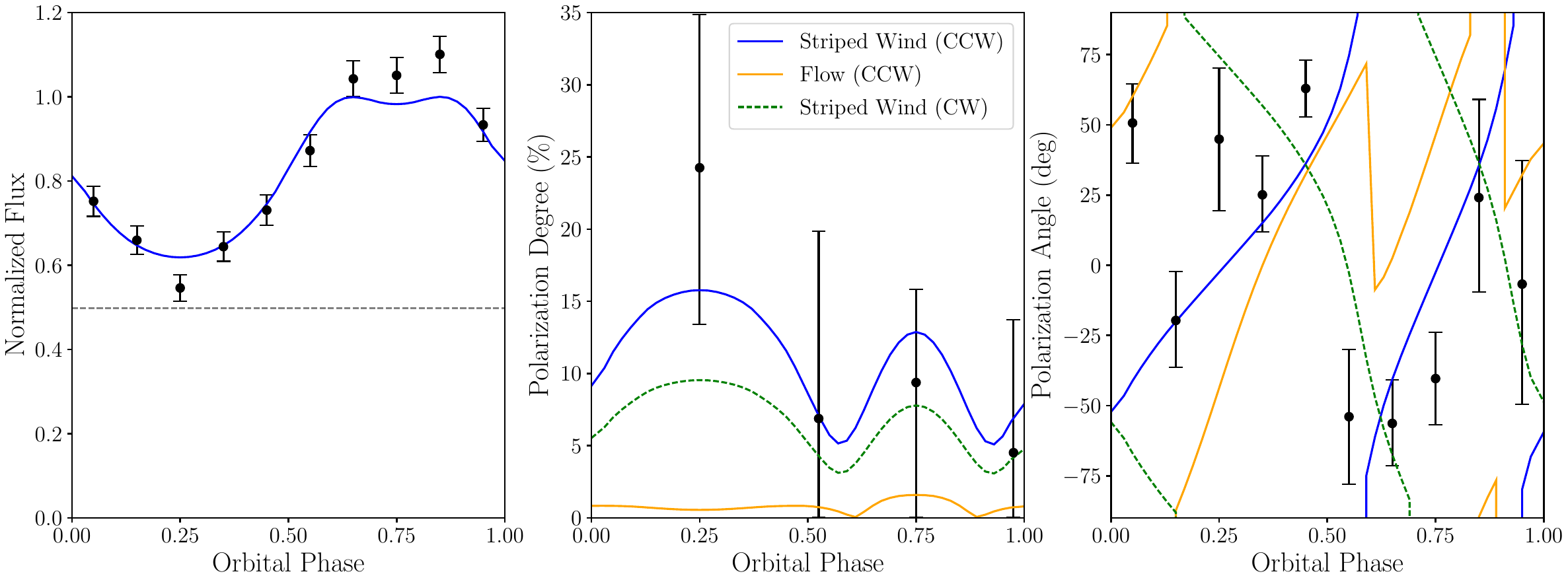}
     \caption{\textit{Left}: The measured {\it IXPE} light curve for J1723 {folded on its orbital period.} The dashed line shows the estimated level of the non-IBS {unmodulated} flux. \textit{Middle and right}: The best fitting counterclockwise (CCW; blue) striped wind model polarization pattern from our unbinned analysis described in Sec.~\ref{sec:Model}, as compared to the deprecated clockwise (CW) and CCW flow models.}
         \label{fig:ixpe_lc}
\end{figure*}
The {\it IXPE} satellite observed J1723 from 21 August to 20 September, 2024 (ObsID: 03006799) for a total exposure of 1.25 Ms in each detector. We process the data using the neural network (NN) based track reconstruction method of \citet{peirson2021deep} including the background weighting method of \citet{dinsmore2025polarization} to obtain enhanced event weights. We remove excess blurring, caused by boom drift over an {\it IXPE} orbit, using the \texttt{ixpeboomdriftcorr} code.

We analyze the data using the \texttt{LeakageLib} software \citep{dinsmore2025leakagelib}, which estimates the source flux and polarization parameters in an unbinned analysis by maximizing the likelihood
\begin{equation}
    L = \prod_{j} \sum_{s} p_{s,j}   P_s(\bm r_j | \psi_j)P_s(\phi_j)P_s(\psi_j | \phi_j),
    \label{eqn:like}
\end{equation}
where $\bm r_j$ and $\phi_j$ are the position in the detector and orbital phase of event $j$, while $\psi_j$ is the electric vector position angle (EVPA) of event $j$. $P_s$ are the probability distribution functions (PDF) of these variables for source $s$. $p_{s,j}$ is the probability of event $j$ originating from source $s$ \citep[see][for details]{dinsmore2025polarization}. For J1723, the EVPA PDF is a function of orbital phase because the flux and expected EVPA are phase-dependent. The factors in Eq.~\ref{eqn:like} act as particle, spatial, temporal, and polarization weights, respectively. Proper use of these weights improves on the standard \texttt{PCUBE} polarization analysis \citep{2015APh....68...45K, 2022SoftX..1901194B}, which {was used by \citet{2025arXiv250905240N} and} did not deliver {particularly constraining} results for these data. {When performing our own \texttt{PCUBE} analysis, we find consistent results with \citet{2025arXiv250905240N}}. \texttt{LeakageLib}'s unbinned maximum likelihood fit method is particularly valuable for sources with modulated EVPA. In a binned analysis, rapid modulations depolarize the signal in individual bins, leading to reduced significances. Our method avoids this problem.

We extract X-ray events in the 2-8 keV {\it IXPE} band. We cut background flares due to excess solar or particle activity by binning the exposure into 800 s intervals and removing intervals with an excess of 600 counts across all three detectors in a region that excludes the central 130$''$ radius around the source. This procedure removes 171 ks of the 1.25 Ms exposure in each detector, leaving behind 81\% of the total 2-8 keV events.

For each detector, we extract photons from a 130$''$ aperture centered on the source.
The large aperture allows for the inclusion of a greater number of source photons. The sky-calibrated {\it IXPE} point spread function (PSF) measured in \cite{dinsmore2024polarization}, which specifies the form of $P_s(\bm r_j | \psi_j)$, de-weights the background counts in the outer regions of the aperture. 

The data prefer a polarized photon background model, with individual polarization degrees (PDs) and EVPAs for each detector. Differently polarized backgrounds have been seen in other {\it IXPE} observations \citep[e.g.~3C 58,][]{2025A&A...699A..33B} and may be due to weak solar flares left in the data. We also see non-uniform particle track position angles, possibly due to secondary cosmic rays emitted from the spacecraft structure. Lacking the data for a full study of this non-uniformity, we model the particle background as a polarized component with constant modulation factor. We fit for the particle and photon background component polarizations and fluxes individually in each detector in an annulus around the source aperture with inner radius 130$''$ and outer radius 260$''$. We subsequently use parameters fixed to this background model in our fit to the source flux and polarization.

The source flux is modeled as point-like, distributed by the individual detector PSFs. {We fold the X-ray counts over the binary orbital period of J1723 using the ephemeris obtained by \citet{2013ApJ...776...20C}.} As discussed in the next section, the data show orbital modulation in addition to a non-varying component, {to which we refer hereafter as the unmodulated component}. The modulated component follows the model light curve in Fig.~\ref{fig:ixpe_lc}. The model specifies $P_s(\phi_j)$. The model polarizations of both the modulated and {unmodulated} components specify the EVPA PDF $P_s(\psi_j | \phi_j)$. The polarization's orbital dependence is shown in Fig.~\ref{fig:ixpe_lc} and described in further detail below.

\section{Model Comparison}
\label{sec:Model}

\begin{figure}
    \centering
    \includegraphics[width=\columnwidth]{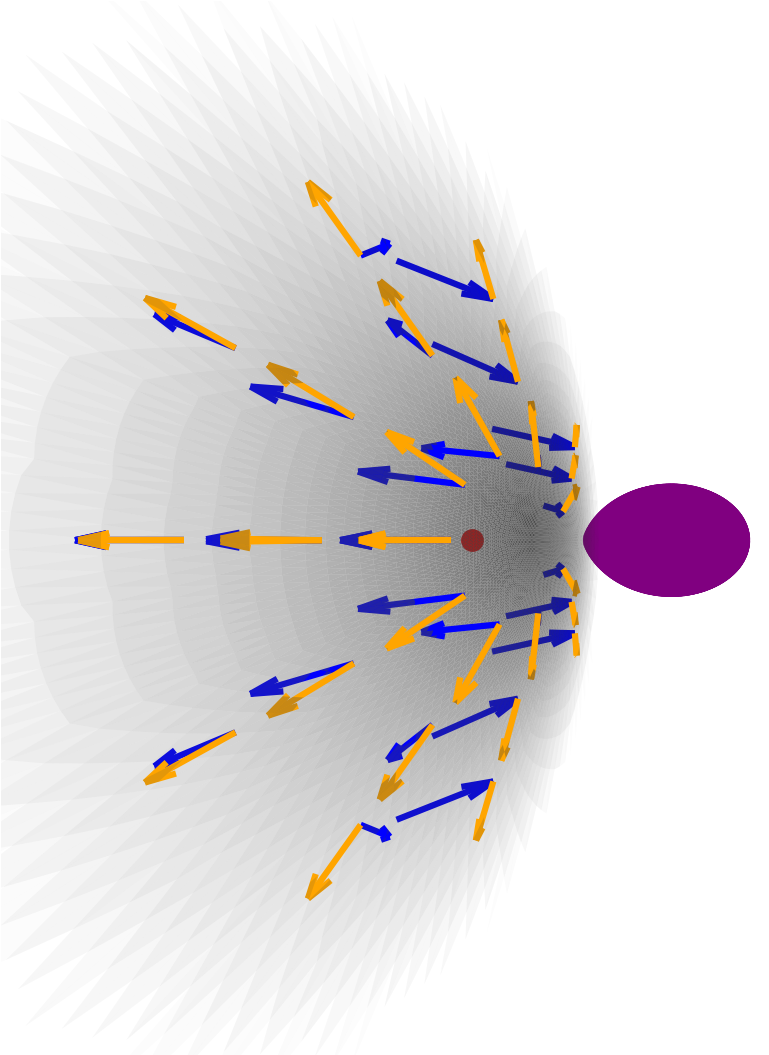}
    \caption{A 3D visualization of the IBS with the striped wind (blue) and flow (orange) magnetic field lines. The pulsar (red dot) and companion are also shown.}
    \label{fig:IBS_model}
\end{figure}

\begin{table*}
\setlength{\tabcolsep}{6pt}
   \hspace{-1.75cm}
\begin{tabular}{lccccccccc}
 
 \hline\hline
 Model & Rotation &{Unmodulated} Component &$\Pi_{\mathrm{IBS}}$ (\%) & $\Psi_{\mathrm{IBS}}$ ($^\circ$)&$\Pi_{\mathrm{IBS}, 99}$ (\%) &  $q_0$ & $u_0$ &$\Delta \mathrm{AIC}$\\ \hline 
 \textbf{Striped wind}&  \textbf{CCW} & &\textbf{$\bf{36^{+16}_{-15}}$} & $\bf{-1^{+16}_{-16}}$ & $\bf{<73}$ &\textbf{0} & \textbf{0}& \textbf{0}\\

 && Constant &$37^{+17}_{-16}$ & $-1^{+17}_{-17}$ & $<76$& $0.00^{+0.12}_{-0.12}$ & $0.00^{+0.12}_{-0.12}$& $-4$\\
  && Rotating PA &$37^{+17}_{-16}$ & $-0^{+17}_{-17}$ &$<76$& $0.00^{+0.12}_{-0.12}$ & $0.00^{+0.12}_{-0.12}$ & $-4$\\
\hline
 Striped wind &  CW & &$25^{+14}_{-12}$ & $88^{+28}_{-28}$ & $<44$&0 & 0& $-3$\\
\hline
Flow 
 &CCW& &$9^{+5}_{-4}$ &  $-87^{+59}_{-59}$ & $<22$&0 & 0& $-4$ \\

  \hline
  
Flow 
 &CW&  &$9^{+5}_{-4}$ &  $-14^{+58}_{-58}$ & $<16$&0 & 0& $-4$ \\

 \hline\hline
\end{tabular}
\caption{Best-fit polarization models for J1723. We compare two IBS polarization models of \citet{2023IBSPolarization} for the two orbit parities. We also check unpolarized and polarized models for the {unmodulated} flux component. 68\% confidence intervals are reported, and $\Pi_\mathrm{IBS,99}$ represents the 99\% upper limit. $\Delta\textrm{AIC}$ is reported in comparison to a fully unpolarized model with $\Pi_{\rm IBS}=0$, $q_0=0$, and $u_0=0$. Statistically preferred models have larger $\Delta$AIC values.}
\label{table:2}
\end{table*}

We model the emission using the \texttt{ICARUS IBS} code \citep{2016ApJ...828....7R, 2019ApJ...879...73K}. \citet{2023IBSPolarization} posit two IBS polarization models: 1) a ``striped wind" model in which the toroidal pulsar wind magnetic field is shock-compressed and imprinted on the IBS and 2) a ``flow" model in which the post shock field is stretched along the shocked IBS flow, tangent to the shock surface. These two models are visualized in Fig.~\ref{fig:IBS_model}. 3D particle-in-cell simulations of striped winds interacting with a companion wind find structure very similar to the striped wind model \citep{2025ApJ...991...98S}. While these simulations are performed for the black widow geometry in which the shock wraps around the companion, the fundamental geometric features which govern the PD and EVPA sweeps such as the toroidal upstream magnetic field will be the same in both the redback and black widow geometries. The overall PD level predicted by \citet{2025ApJ...991...98S} varies depending on the stripe-averaged magnetic field and the level of downstream turbulence relative to this residual field.
\subsection{Optically Derived Inclination}
\begin{figure}
   \centering

   \includegraphics[width=\linewidth]{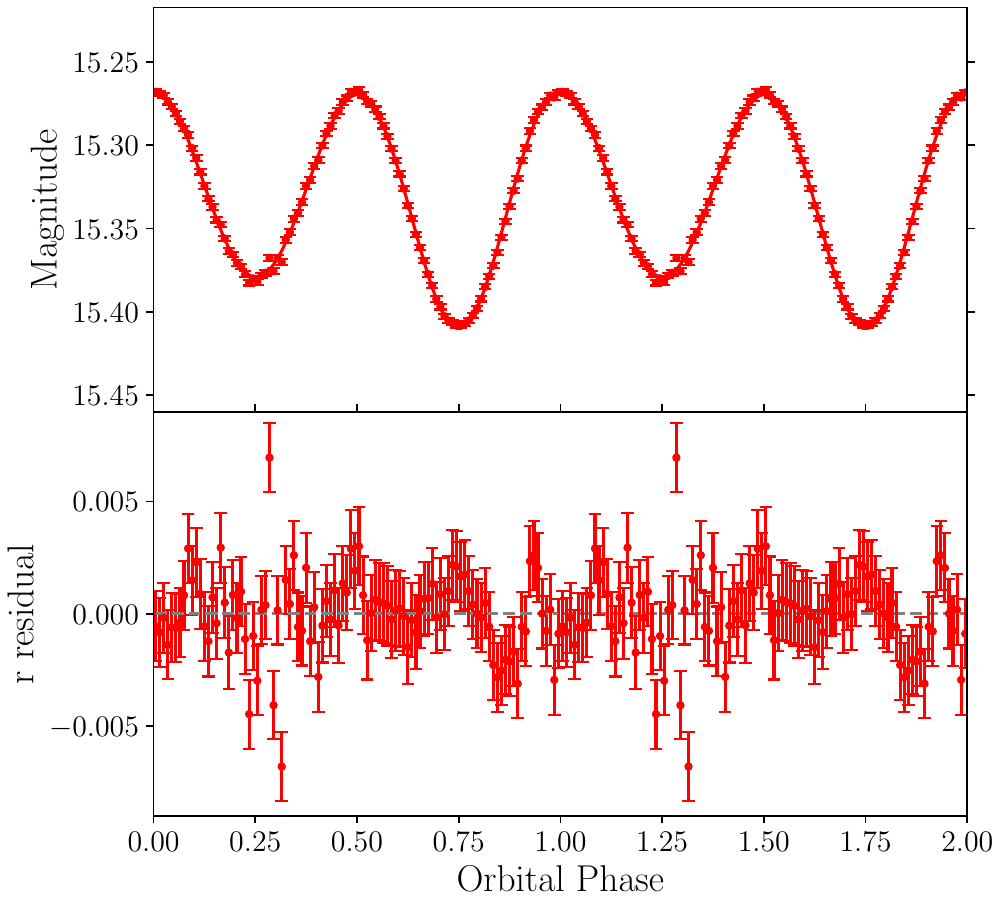}
     \caption{Quiescent r-band K2 light curve (dominated by ellipsoidal modulation) and residuals to the best fit model. Errors are inflated from their originally statistical values to account for rapid modulation, plausibly due to low level companion flare activity.}
         \label{fig:rlc}
\end{figure}

The IBS X-ray light curve and polarization depend on the binary viewing inclination $i$, which can be measured from fits to the companion's optical light curve \citep{2013ApJ...769..108B, 2018ApJ...859...54L, 2020ApJ...892..101K, 2021ApJ...908L..46R, 2024ApJ...974..315S}.  For J1723, the companion light curves are dominated by ellipsoidal modulation, but also show asymmetries that can be interpreted as a hot spot. The companion's hot spot appears to vary in phase, possibly due to asynchronous rotation \citep{2016ApJ...833L..12V} or non-radial oscillations \citep{2021MNRAS.508.3812N}. The low-mass companion is convective and rapidly rotating, so these hot spots can be produced by a magnetic dynamo, which may also drive flaring activity seen at many epochs.
A Kepler K2 light curve is available on the archive \citep[target ID=236020326, campaigns 111/112, covering 121 orbits,][]{2014PASP..126..398H}. To isolate the quiescent, flare-free behavior, we create an average light curve from a low variability portion---orbits $90-100$. The light curve is fit with the \texttt{ICARUS} code \citep{2013ApJ...769..108B}. {We fix the Roche lobe filling factor at $f_c=0.999$, the radial velocity amplitude at $K=143$ km s$^{-1}$ \citep[the central value in the range estimated by][]{2013ApJ...776...20C}, the distance to the source at 0.8 kpc \citep{2013ApJ...776...20C}, and the extinction at $A_V=0.12$ based on the dust map of \citet{2024A&A...685A..82E} at this distance.} 
We fit for $i$, base stellar temperature $T_N$, a calibration offset from the Kepler magnitude to the SDSS r-band $\Delta m_r$, and a surface hot spot. Interestingly, there is no evidence for pulsar heating with a 99\% confidence upper limit of $L_H<3 \times 10^{-5}\, {\dot E}_{\rm PSR}$ on the heating flux.

Fig.~\ref{fig:rlc} shows the best-fit model alongside the observed light curve. The residuals show milli-mag modulations, likely due to unresolved minor flaring present even in near-quiescence. These lead to a substantial $\chi^2$. We treat these unmodeled fluctuations as an error, and increase the statistical errors by $0.0015\,$mag. These inflated errors are shown in Fig.~\ref{fig:rlc} rather than the extremely small statistical errors. 
With this inflation, we find $\chi^2/\mathrm{DoF}=1.3$. For our purposes, the most important parameter is the orbital inclination. We obtain $i=40.3^{+0.2}_{-0.1}$$\,^\circ$. This value is fully consistent with that found in \citet{2016ApJ...833L..12V} and we adopt it for IBS modeling. 

\subsection{Modeling the Intrabinary Shock}

To model the IBS, we use the ratio of the companion wind to pulsar wind momentum fluxes $\beta=\dot{M}v_wc/\dot{E}_\mathrm{PSR}=5$, which determines the phase separation of the two flux peaks at a given $i$. The model can also produce asymmetry in the IBS peaks, parameterized by the ratio of orbital speed to that of the companion wind $v_{\rm orb}/v_w$. For $\beta>1$, the leading peak is expected to be brighter. In our {\it IXPE} light curve, the trailing peak appears brighter, but in higher significance NuSTAR and XMM J1723 light curves \citep{2025ApJ...984..146S}, the peaks are symmetric, so we set the velocity ratio to 0 in our models. Following \citet{2019ApJ...879...73K, 2023IBSPolarization}, we assume that the bulk Lorentz factor of the post-shock flow (assumed parallel to the contact discontinuity) increases with length along the IBS $s$ as 
\begin{equation}
    \Gamma_{\rm Bulk}(s)=\Gamma_{\rm nose}\left(1+k\frac{s}{r_0}\right),
\end{equation}
where $r_0$ is the distance from the pulsar to the IBS nose. We choose $\Gamma_{\rm nose}=1.1$ and $k=0.08$ based on previous redback IBS modeling \citep{2024ApJ...974..315S}. We assume that the shock-compressed pulsar wind magnetic field dominates in the radiation zone. We take $B_{\mathrm{IBS}}= 3 B_{\rm LC} (r_{\rm LC}/r_0)=38$\,G at the nose, where $B_{\rm LC}$ is the magnetic field at the light cylinder and $r_{\rm LC}$ is the light cylinder radius. See \citet{2023IBSPolarization} for further details on the treatment of the magnetic field geometry in the two models. We select particle spectral index $p=0.7$ to match the observed IBS X-ray spectrum \citep{2025ApJ...984..146S}.

In addition to the orbit-modulated emission from the IBS, there is evidence for a softer phase-independent non-thermal flux \citep{2025ApJ...984..146S}, which we refer to as the {unmodulated} component. The source of this flux is unknown, but may come from particles stalled at the IBS nose, an unresolved PWN, or the magnetosphere. We find no significant signal when checking for spin modulation of this flux, making magnetospheric emission unlikely. IBS nose and PWN emission remain plausible. Since these components would also likely be synchrotron, we investigate possible polarized {unmodulated} models: 1) a ``constant"  model where the polarization does not vary, and 2) a ``rotating PA" model where the polarization sweeps, assuming the magnetic field rotates with the binary in the orbital plane, with a constant PD. 

The phase of the EVPA sweep and the PD amplitude reveal the orientation of the orbital angular momentum on the plane of the sky and the turbulence in the post-shock field. When quoting our results, we report the EVPA $\Psi_\mathrm{IBS}$ at phase of pulsar superior conjunction $\phi_b=0.25$. At this phase, the EVPA is equal to the orbital angular momentum direction projected on the sky in both models. We quote the maximum IBS PD $\Pi_\mathrm{IBS}$, which is at $\phi_b=0.25$ in the striped wind model and $\phi_b=0.75$ in the flow model. Magnetic turbulence suppresses $\Pi_\mathrm{IBS}$ below the maximum uniform field value {expected from classic synchrotron theory \citep{1979rpa..book.....R}} $\Pi_{\rm max} = (p+1)/(p+7/3)\times100\% = 56\%$ {with $p=0.7$}. Note that the direction of the EVPA sweep depends on whether the orbital angular momentum is into or out of the plane of the sky, giving either a clockwise (CW) or counterclockwise (CCW) sweep. We parameterize the {unmodulated} component with Stokes coefficients $q_0$ and $u_0$. For the ``rotating PA'' {unmodulated} model, $q_0$ and $u_0$ are reported at $\phi_b=0.25$, and the rotation direction is chosen to be the same as that of the IBS. 

The fit results are presented in Table \ref{table:2}. We report the Akaike information criterion (AIC) of each model as compared with a completely unpolarized model ($\Pi_{\rm IBS}=0$, $q_0=0$, and $u_0=0$), with positive $\Delta$AIC indicating preference over the unpolarized model. The best-fitting model is the CCW striped wind. While the {best-fit} CCW striped wind {achieves a greater maximum likelihood} than the completely unpolarized scenario, the maximum likelihood improvement is insufficient to overcome the AIC penalty from the additional parameters, so $\Delta \mathrm{AIC}=0$. With no polarized model reaching high significance, we also give the 99\% confidence upper limit on the IBS component PD $\Pi_{\rm IBS,99}$. As expected, these upper limits are stronger for deprecated models; inaccurate models should generally yield reduced PDs because they average over unmodeled EVPA sweeps, reducing polarization.

The best-fit polarization of the {unmodulated} component is consistent with zero in all cases, and fitting for {unmodulated} polarization decreases the AIC. Therefore, in Table \ref{table:2}, we only show results for a polarized {unmodulated} model in the CCW striped wind case, to illustrate this decrement.

We show the best-fitting CCW striped wind, CW striped wind, and CCW flow models alongside the binned light curves in Fig.~\ref{fig:ixpe_lc}. Note that the PDs of Fig.~\ref{fig:ixpe_lc} are lower than the $\Pi_{\mathrm{IBS}}$ central value reported in Table \ref{table:2}, as the total flux is diluted by the unpolarized {unmodulated} component.

\section{Discussion and Conclusion}
\label{sec:conclusion}

We conclude that the CCW striped wind model is the best of the polarized models. This suggests CCW rotation of the binary on the sky. The results are also most consistent with the natural striped wind picture, in which the plasma radiates in the intrabinary shock-compressed toroidal pulsar wind magnetic field \citep{2023IBSPolarization, 2025ApJ...991...98S}. This contrasts with the flow model in which the IBS post-shock flow would stretch the field lines along the contact discontinuity. For the striped wind model $\Pi_{\rm IBS}=36^{+16}_{-15}\%$, which is safely less than $\Pi_\mathrm{max} = 56\%$ for $p=0.7$. Our measured model PD is greater than 0 by $\sim2.4\sigma$, but unfortunately this model is not yet statistically distinguished from the unpolarized null model.


\citet{2025ApJ...991...98S} perform a suite of black widow simulations in which the pulsar wind is treated as a plane-parallel striped wind. These simulations are parameterized by $\alpha_{\rm pw}=2\langle B\rangle_\lambda/(B_0+\langle B\rangle_\lambda)$ (not to be confused with the pulsar magnetic obliquity also usually denoted by $\alpha$ and here by $\alpha_B$) where $\langle B\rangle_\lambda$ is the average upstream magnetic field across one striped wind wavelength and $B_0$ is the upstream field. In the pulsar wind, the value of $\alpha_{\rm pw}$ varies from 0 to 1, away from the spin equator, over an angle comparable to the pulsar spin obliquity \citep{1999A&A...349.1017B}.
In black widows, the shock often subtends an angle smaller than the pulsar spin obliquity, so $\alpha_{\rm pw}\approx0$ across most of the shock in these spin-aligned systems \citep{2025MNRAS.tmp..266C, 2025ApJ...991...98S}. In redbacks, where the shock captures a large solid angle of the pulsar wind, $\alpha_{\rm pw}$ varies with latitude. Following \cite{1999A&A...349.1017B}, the magnetic field in the striped wind is
\begin{equation}
\label{eq:mag}
    B=B_0\,D\left(\sin i_p\sin\left(\frac{r}{r_{\mathrm{LC}}}\right)\sin\alpha_B+\cos i_p\cos\alpha_B\right),
\end{equation}
where $i_p$ here is the angle from the pulsar spin axis, $r$ is the distance from the pulsar, and  \citep{1999A&A...349.1017B}
\begin{equation}
    D(x)=\begin{cases}
    -1, & x<0\\
    0, & x=0\\
    1, & x>0
    \end{cases}.
\end{equation}
Averaging Eq.~\ref{eq:mag} over the striped wind wavelength $\lambda=2\pi r_{\rm LC}$ gives $\langle B\rangle_\lambda$ and thus $\alpha_{\rm pw}$ for each position at the IBS.

\begin{figure}
   \centering

   \includegraphics[width=\linewidth]{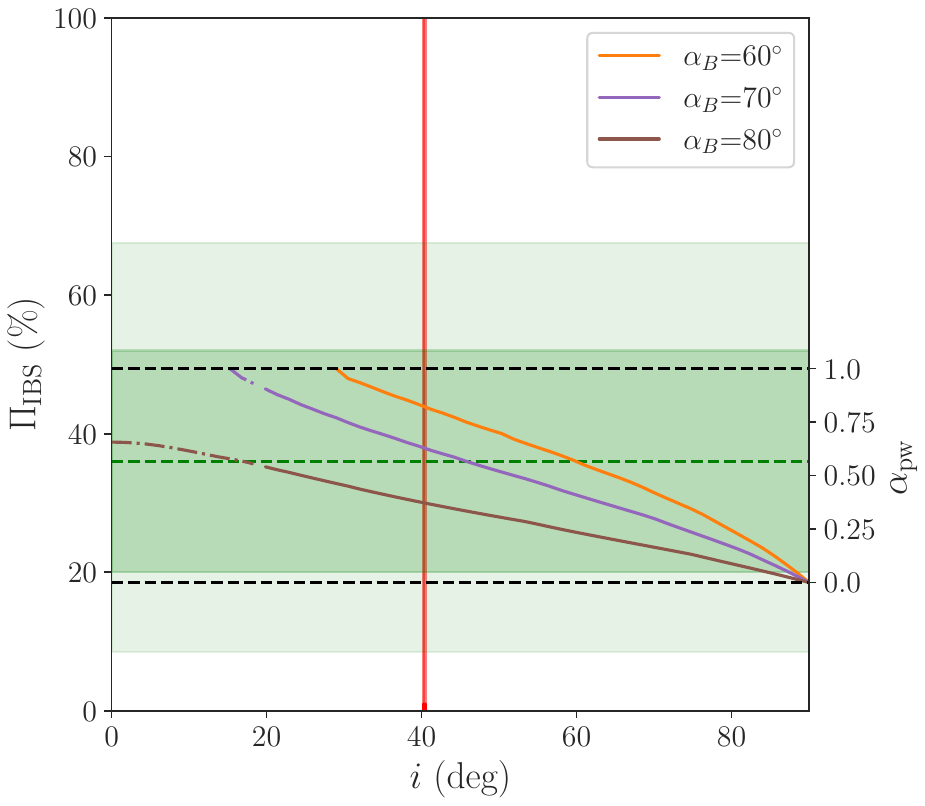}
         \caption{Shaded bands show the measured J1723 1$\sigma$ and 2$\sigma$ confidence regions for $\Pi_{\rm IBS}$. The $\Pi_{\rm IBS}$ values between the black dashed lines map to $\alpha_{\rm pw}$ values shown on the right axis using the linear version of eq. 10 from \citet{2025ApJ...991...98S}, for $i=60^\circ$. We show the phase-averaged-flux-weighted mean $\alpha_{\rm pw}$ across our model IBS as a function of binary viewing inclination $i$ for different magnetic obliquity angles $\alpha_B$ and mark our light curve-measured $i=40.3^\circ$. At $i<20^\circ$, the IBS emission is  unlikely to be detected (see text), so we continue the curves with dashed-dotted lines.}
         \label{fig:alpha}
\end{figure}

The simulations sample $\alpha_{\rm pw}$ ranging from 0 to 0.5 and estimate PDs from $\sim15\%-30\%$ at flux maximum. 
\citet{2025ApJ...991...98S} estimate a linear trend between $\alpha_{\rm pw}$ and PD at flux maximum.  
The PD at flux maximum (at $\phi_b=0.75$ in redbacks) is $0.81$ times its value at PD maximum (at $\phi_b=0.25$ in redbacks) in our model.
Although approximate  \citep[since the][simulations assume a plane-parallel striped wind incident on a black widow companion]{2025ApJ...991...98S}, we estimate $\alpha_{\rm pw}$ by substituting $\Pi_{\rm IBS}\times0.81=29^{+13}_{-12}\%$ into the trend from \citet{2025ApJ...991...98S} for $i=60^\circ$ (since this was the lowest inclination tested) and obtain $\alpha_{\rm pw}\approx0.6^{+0.4}_{-0.5}$.  Fig.~\ref{fig:alpha} shows our measured 1$\sigma$ and 2$\sigma$ confidence ranges for $\Pi_{\rm IBS}$ at $\phi_b=0.25$. On the right axis, we show the corresponding estimated $\alpha_{\rm pw}$ values from the aforementioned mapping. From our model, we plot the mean IBS $\alpha_{\rm pw}$ weighted by the phase-averaged flux of each zone as a function of binary inclination $i$ for different choices of $\alpha_B$. At $i<20^\circ$, the IBS emission is suppressed by a factor of 3 from its value at $i=90^\circ$, so we are unlikely to detect it. Consequently, we continue the $\alpha_{\rm pw}$ curves with dashed-dotted lines. Note that
at low $i$, large $\alpha_{\rm pw}$ values are expected, depending on obliquity $\alpha_B$. $\alpha_{\rm pw}\approx0.6$ and $i=40.3^\circ$ (denoted by the red band) would suggest obliquity $\alpha_B\approx70^\circ$. Additionally, $\alpha_{\rm pw}\approx0.6$ means that the stripe-averaged field and thus the post-shock residual field would be $\sim40\%$ its upstream value. This is consistent with the magnetic field inferred from spectral analyses of J1723 \citep{2025ApJ...984..146S}. Thus, our observed  $\Pi_{\rm IBS}$ is well consistent with expectations from IBS models.

This model cannot yet be significantly distinguished from the unpolarized null model and requires more data to make a detection. The high formal $\Pi_{\mathrm{IBS},99}$ bound is for the pulsed IBS component which accounts for only half the total flux. When including the {unmodulated} component, our total PD upper limit is $\Pi_{99}\lesssim36\%$ at PD maximum. An additional 1 Ms {\it IXPE} exposure could decrease the $1\sigma$ error for the total PD to $\sim11\%$, and could push the $\Pi_{\mathrm{IBS}}$ measurement to $3\sigma$. Since $\Psi_{\rm IBS}$ gives the projection of the binary angular momentum axis on the plane of the sky, comparison with radio pulsar polarization results, which provide a measurement of the projected spin axis, can test spin-orbit alignment (as might be expected in spin-up scenarios).   

Additional numerical simulations of IBSs in the redback geometry may provide other polarization models and an improved basis for interpretation of these results. Certainly, a high significance polarization measurement with improved phase resolution could provide powerful tests of the shock physics and reconnection-driven particle acceleration, as the orbit provides varying views of the oblique shock system.

During the writing of this manuscript, \citet{2025arXiv250905240N} posted an analysis of the J1723 data, using the more basic \texttt{PCUBE} method. This provided only upper limits on the {\it total average} PD of $\Pi_{99}\lesssim50\%$ in two phase bins. For the IBS component alone, this corresponds to an average PD of $\lesssim 100\%$ (assuming an unpolarized {unmodulated component}), and an effectively unconstraining maximum PD of $\Pi_{\rm IBS,99}\lesssim 130\%$ (since the maximum PD is $\sim$ 30\% larger than the average in the best-fitting model). They compare these measurements to simple geometric IBS magnetic field models, one of which is reminiscent of our flow model. In the present paper, using the NN-derived track reconstruction and PSF-fitting methods of \texttt{LeakageLib}, we place stronger PD limits and track the phase-resolved polarization sweep. Furthermore, unbinned phase-resolved model fitting enables direct comparisons between polarization models and helps us estimate the overall PD level preferred by each model. \citet{2025arXiv250905240N} also perform a model-independent analysis to estimate the EVPA rotation direction and report a preference for CW rotation. We are able to directly compare the rotation direction in our {polarized} models and in contrast find a preference for CCW rotation by $\Delta\mathrm{AIC}=3$ under the {polarized} striped wind model. An additional 1 Ms exposure analyzed with standard \texttt{PCUBE}-type methods would approximately match the precision of the results in this paper; however, analyzing a longer exposure with our advanced extraction and analysis techniques would improve upon the present measurement.


Polarization observations now probe the magnetic geometry of the pulsar wind across different regimes. {\it IXPE} observations have shown significant X-ray polarization in transitional MSPs \citep{2025ApJ...987L..19B}, where the pulsar wind may shock against the inner edge of an accretion disk at a few $r_{\rm LC}$. Spider IBSs probe the wind at $r\sim 0.1-1\,R_\odot \approx 10^3-10^4\, r_{\rm LC}$ and include zones away from the orbital plane, so our observations provide a complimentary pulsar wind probe. Our spider IBS X-ray polarization study represents the most constraining probe to date on the shocked pulsar wind magnetic field structure at this scale. Our measurements, enabled by PSF fitting and enhanced background modeling \citep{dinsmore2025polarization}, prefer shock-imprinted striped wind models to models in which the post-shock magnetic field is stretched with the bulk flow along the IBS. While stronger constraints would require a substantially longer {\it IXPE} exposure or a next-generation facility, our IBS polarization measurements along with those of PWNe observationally trace the multi-scale properties of pulsar winds.

\section*{Acknowledgements}
The authors are grateful to Roger Blandford, Lorenzo Sironi, and Maya Beleznay for useful discussions. This work was supported in parts by contract NNM17AA26C from the MSFC to Stanford in support of the {\it IXPE} project and NASA grant 80NSSC25K7096. A.S.~acknowledges the support of the National Science Foundation Graduate Research Fellowship Program and a Giddings Fellowship to the Kavli Institute for Particle Astrophysics and Cosmology at Stanford University.
\bibliography{refs}{}
\bibliographystyle{aasjournal}

\end{document}